%% file: interop-pr.tex
\documentclass[5p,times]{elsarticle}

\usepackage{booktabs} 

\graphicspath{{fig/}}

\usepackage[caption=true,font=footnotesize]{subfig}
\usepackage[font=small]{caption}

\usepackage[breaklinks,hidelinks]{hyperref}

\usepackage{listings}

\usepackage{enumitem}

\usepackage{newtxtext,newtxmath}

\makeatletter
\def\ps@pprintTitle{%
 \let\@oddhead\@empty
 \let\@evenhead\@empty
 \def\@oddfoot{}%
 \let\@evenfoot\@oddfoot}
\makeatother

\begin{document}

\begin{frontmatter}

\title{Integrating Blocking and Non-Blocking MPI Primitives with\\Task-Based Programming Models}

\author[BSC]{Kevin Sala\fnref{ORCID}}
\ead{kevin.sala@bsc.es}
\address[BSC]{Barcelona Supercomputing Center (BSC)}
\fntext[ORCID]{ORCIDs: 0000-\{0001-8233-1185, 0001-5181-7545, 0002-0558-7600, 0002-3575-4617, 0002-3580-9630, 0002-7489-4727\}}

\author[BSC]{Xavier Teruel\fnref{ORCID}}
\ead{xavier.teruel@bsc.es}

\author[BSC]{Josep M. Perez\fnref{ORCID}}
\ead{josep.m.perez@bsc.es}

\author[BSC]{Antonio J. Pe\~na\fnref{ORCID}}
\ead{antonio.pena@bsc.es}

\author[BSC]{Vicen\c{c} Beltran\fnref{ORCID}}
\ead{vbeltran@bsc.es}

\author[BSC]{Jesus Labarta\fnref{ORCID}}
\ead{jesus.labarta@bsc.es}

\input{tex/abstract}



\end{frontmatter}

\input{tex/introduction}
\input{tex/background}
\input{tex/related}
\input{tex/apis}
\input{tex/mpi}
\input{tex/tampi}

\input{tex/evaluation}
\input{tex/standardization}
\input{tex/conclusion}

\input{tex/acknowledgments}

\bibliographystyle{elsarticle-num}
\bibliography{interop-pr}

\end{document}

%% file: tex/abstract.tex
\begin{abstract}

In this paper we present the Task-Aware MPI library (TAMPI) that integrates both blocking and non-blocking MPI primitives with task-based programming models.
The TAMPI library leverages two new runtime APIs to improve both programmability and performance of hybrid applications.
The first API allows to pause and resume the execution of a task depending on external events. This API is used to improve the interoperability between blocking MPI communication primitives and tasks.  When an MPI operation executed inside a task blocks, the task running is paused so that the runtime system can schedule a new task on the core that became idle.
Once the blocked MPI operation is completed, the paused task is put again on the runtime system's ready queue, so eventually it will be scheduled again and its execution will be resumed.

The second API defers the release of dependencies associated with a task completion until some external events are fulfilled.
This API is composed only of two functions, one to bind external events to a running task and another function to notify about the completion of external events previously bound.
TAMPI leverages this API to bind non-blocking MPI operations with tasks, deferring the release of their task dependencies until both task execution and all its bound MPI operations are completed.

Our experiments reveal that the enhanced features of TAMPI not only simplify the development of hybrid MPI+OpenMP applications that use blocking or non-blocking MPI primitives but they also naturally overlap computation and communication phases, which improves application performance and scalability by removing artificial dependencies across communication tasks.

\end{abstract}

%% file: tex/introduction.tex
\section{Introduction}
\label{sec:intro}

Current near-term and mid-term high-performance computing (HPC) architecture trends suggest that the first generation of exascale computing systems will consist of distributed memory nodes, where each node is powerful and contains a large number of compute cores.
A well-established practice in the HPC community is to develop hybrid applications combining APIs such as MPI and OpenMP, which are specialized in exploiting inter-node and intra-node parallelism, respectively.
Although MPI and OpenMP were not originally designed to be used together, these have evolved to provide some interoperability support. 
However, this minimal support heavily determines how both models can be
safely combined to develop hybrid applications, posing
performance implications.

The MPI Standard guarantees that point--to--point communications
among two ranks are
always ordered as long as these leverage the
same tag and communicator.
However, when multiple threads communicate simultaneously, the operations are logically concurrent and hence these threads can receive them in any order.
To avoid ordering problems on hybrid applications, in practice MPI
communications are usually restricted to sequential parts of the
application (what is known as MPI's thread funneled mode), while most computations are performed in parallel. 
This results in a common pattern that interleaves parallel computation phases (fork--join) with sequential communication phases.
This is the easiest and most common way to combine both programming
models, but it is not free of drawbacks.
On the one hand, it is not easy to overlap computation and communication phases; on the other hand, both inter-node and intra-node parallelism may be potentially hindered due to the strict synchronization enforced among computation phases and across nodes.
Hybrid applications may be restructured to manually overlap computation and communication phases of the algorithm using asynchronous communication primitives and techniques such as double-buffering.
However, these techniques require complex modifications of the code that, depending on the application complexity, are not even feasible.

An easy way to solve the previous issues would be to parallelize both computation and communication phases using tasks, relying on task dependencies to deal with inter-node and intra-node synchronizations.
However, this approach cannot be efficiently implemented with current MPI and OpenMP specifications.
MPI provides the {\sf MPI\_THREAD\_MULTIPLE} mode that supports the concurrent invocation of MPI calls from multiple threads, but this is not sufficient to efficiently support task-based programming models such as OpenMP.
The main issue is that tasks are not aware of the synchronous MPI primitives, which might block not only the task but also the underlying hardware thread that runs it.
Even if the MPI implementation does not rely on busy-waiting to check
for operation completion and the hardware thread becomes idle, the
task runtime has no means to discover that the hardware thread is
available without an explicit notification from the MPI side.  
Without this notification mechanism, if the number of in-flight MPI operations blocked reaches the number
of available hardware threads, the application will hang due to lack of progress. 
With the current MPI Standard, the application developer is the responsible for avoiding this situation.
However, this severely limits the ability of developers to fully benefit from task-based programming models.

In ~\cite{interopmpi}, the authors introduced a new level of threading support called {\sf MPI\_TASK\_\-MULTIPLE} that overcomes most of the limitations involved in mixing tasks and blocking MPI operations. This new level of threading support was implemented in both a native MPI library and the portable Task-Aware MPI (TAMPI) library.
In this paper, we extend the task-based runtime system and the TAMPI library to efficiently support non-blocking MPI primitives inside tasks. 
In this approach, communication tasks bind the release of their dependencies to the completion of the MPI requests that they initiate. This way, other tasks, by declaring the proper dependencies over those buffers, can either consume (read after receive), or reuse them (write after send).
Hence, communication tasks are no longer blocked, which is essential in applications that have many communication tasks or communicate small messages.

In this work, (1) we introduce a generic API to programmatically pause and resume task execution;
(2) we introduce another generic API to programmatically decouple the release of task dependencies from the task finalization;
(3) we propose {\sf MPI\_TASK\_\-MULTIPLE}, a new level of thread support for MPI that leverages the pause/resume API to better support blocking MPI operations inside tasks;
(4) we propose two new functions {\sf TAMPI\_Iwait} and {\sf TAMPI\_Iwaitall} that leverage the external events API to better support non-blocking MPI operations inside tasks;
(5) we implement both proposals inside a portable MPI wrapper library that works with any MPI implementation,
and (6) we provide an in-depth performance and scalability evaluation of our proposals.

The rest of the paper is structured as follows.
Section~\ref{sec:background} provides an introduction to the OmpSs-2 and MPI programming models.
In Section~\ref{sec:related} we review related literature.
In Section~\ref{sec:apis} we present the task pause/resume and external events APIs.
Section~\ref{sec:mpi} describes the deadlock situation when calling blocking MPI primitives within tasks.
Section~\ref{sec:interop_lib} describes the TAMPI library.
We evaluate our work in Section~\ref{sec:evaluation}.
In Section~\ref{sec:standardization} we describe the impact of our proposals to the OpenMP standard.
Finally, Section~\ref{sec:conclusion} provides concluding remarks.

%% file: tex/background.tex
\section{Background}
\label{sec:background}
In this section we provide a brief overview of the OmpSs-2 and MPI
programming models along with the implementations we leverage.

\subsection{OmpSs-2}

OmpSs-2 is the second generation of the OmpSs programming model developed at the
Barcelona Supercomputing Center. It is open source and mainly used as a research platform to
conceive, implement and test new ideas that can be exported to the OpenMP
tasking model. OmpSs-2 (like OpenMP) is based on directives and it enables the
parallelism in a data-flow way.  The developer is in charge of decomposing the
code into \emph{tasks} and identifying their data dependencies. This
information is later used by the source--to--source Mercurium~\cite{mcc}
compiler to generate the corresponding calls to the Nanos6~\cite{nanos6}
runtime API. The runtime library is responsible for scheduling and executing
the annotated tasks, preserving the implied task dependency constraints.
The Nanos6 runtime and the Mercurium compiler are publicly available at~\url{http://pm.bsc.es}.

\subsection{MPI}
MPI is a message-passing standard~\cite{mpi_standard} broadly used by the
HPC community. MPICH is a popular open source MPI
implementation (see \url{http://www.mpich.org}), and its derivatives (such as Intel's, Cray's, or IBM's
MPI) are default in 9 out of the top 10 supercomputers in the current
TOP500 list~\cite{top500}.
In this paper we have used both MPICH 3.2.1 (Nov. 2017) and Intel MPI 2017 Update 4.

%% file: tex/related.tex
\section{Related Work}
\label{sec:related}

Overlapping computation and communication phases is a critical issue that has
already been studied in several contexts. In \cite{beltran01, beltran02}, the
authors developed a threading library for the Cell B.E. processor that
transparently overlaps the computation and communication phases of different
threads running on the same SPU (Synergistic Processor Unit in the Cell B.E.
architecture). When the running threads are about to block on a DMA operation
(which would be equivalent to a wait all or wait any operation in the MPI
interface), the execution of the thread is suspended until the DMA operation
completes. In the meantime, the execution of another thread is resumed on the
SPU to overlap the communication phase of the suspended thread(s) with the
computation phase of the current thread. This work also studies double- and
multi-buffering techniques which also allows overlapping, but are limited to
applications with a regular and static communication patterns, while the
approach based on threads supports irregular applications with a dynamic
communication pattern.

The study of hybrid approaches~\cite{par_pm, Raben, Jost} combining communication
libraries and shared memory programming models has been considered
over the last years both in research and in performance
analysis publications.

By using the {\sf comm\_thread} approach of the hybrid MPI + SMPSs programming
model~\cite{smpss_mpi}, authors allowed to exploit distant parallelism
separated by taskified MPI calls. These tasks were also identified as
\emph{communication tasks} and were executed by an additional thread called
\emph{communication thread}. The runtime's task scheduler could reorder the
execution of communication and computational tasks in such a way that
communication can happen as soon as possible, increasing the parallelism within
and across MPI processes.  That proposal requires changes to the programming
model to allow to identify ahead of time those tasks that have blocking-like
behavior.  In addition, only one thread can execute them, and it must do so in
sequential order. Hence, this solution is suboptimal. In this paper we propose
a runtime-agnostic solution that does not require to pre-classify the work
units, which allows tasks to contain any mixture of computations and
communications, and supports several communications in parallel and out of order.

In the Habanero-C MPI (HCMPI) proposal~\cite{chatterjee2013}, MPI calls 
are tightly integrated with the task dependency system. HCMPI treats
all MPI calls as (asynchronous) tasks, which brings well-known issues inherent
to excessively fine-grained tasking, such as increased scheduling overhead and
load imbalance.
In contrast, our proposal for blocking MPI primitives is orthogonal to the dependency system and is
specially well suited to parallelize legacy and library codes, since it does
not require to taskify every MPI operation, hence resulting in a more natural
and flexible approach.
Nonetheless, TAMPI's support for non-blocking primitives is slightly connected to task dependencies, but not integrated into the dependency system.
Through a generic API, it modifies the conditions that tasks have to meet to release their dependencies.

%% file: tex/apis.tex
\section{Interoperability between Parallel Runtimes and Blocking\slash
  Non-Blocking Operations}
\label{sec:apis}

This section overviews and proposes solutions to the challenges of interoperating efficiently parallel runtimes with both blocking and non-blocking operations.

For instance, synchronous I/O operations over files may block the thread that invokes them for the duration of the operation.
On an environment with multiple processes competing for CPU time, the time that the thread is blocked may be used by another thread.
However, on hosts dedicated to a single multithreaded HPC job, the core is most likely to remain idle.

In this section we discuss our proposal in the context of OmpSs-2 and MPI.

\subsection{Block and Unblock}

\lstset{language=C++,basicstyle=\linespread{1.15}\footnotesize\sffamily,fontadjust,columns=fullflexible,tabsize=3,showstringspaces=false}
\lstset{emphstyle=\textit}
\lstset{emph=[2]{get_current_blocking_context,block_current_task,unblock_task,get_current_event_counter,increase_current_task_event_counter,decrease_task_event_counter,register_polling_service,polling_service_t,unregister_polling_service}}
\lstset{morecomment=[l][{\linespread{1.15}\bf\footnotesize\sffamily}]{\#}}

To support the efficient execution of blocking-like operations in parallel runtimes, we first propose an API to pause and resume tasks.
It is composed of three functions. The first has the following prototype in C:
\begin{lstlisting}
    void *get_current_blocking_context();
\end{lstlisting} 

This function informs the runtime that the current task is about to enter a pause--resume cycle.
The function configures everything needed to handle one round trip, and returns an opaque pointer to runtime-specific data.
Throughout the rest of this text we call this data a \emph{blocking context}.
A blocking context is valid only for one pause--resume cycle, and
requesting a new context invalidates the currently active one.

The pause and resume operations are requested through the following functions:
\begin{lstlisting}
    void block_current_task(void *blocking_ctx);
    void unblock_task(void *blocking_ctx);
\end{lstlisting} 

On a call to the first function, the runtime suspends the execution of the invoking task.
The parameter must be the current blocking context of the invoking task.
The second function indicates that the task associated to the blocking context can be resumed.
This function can be called by any thread over a valid blocking context.

\begin{figure*}
 \centering
 \subfloat[Code that performs the blocking operation]{%
 \label{code:pause-pattern}%
\begin{lstlisting}[numbers=left,numberstyle=\tiny,numbersep=5pt]
async_handler = start_async_op(...);
void *blocking_ctx = get_current_blocking_context();
associate(async_handler, blocking_ctx);
block_current_task(blocking_ctx);
\end{lstlisting}%
}
\hfil
\subfloat[Body of the code that handles the unblocking of the operation]{%
 \label{code:resume-pattern}%
\begin{lstlisting}[numbers=left,numberstyle=\tiny,numbersep=5pt]
async_handler = wait_until_one_async_op_finishes();
void *blocking_ctx = get_assigned_blocking_context(async_handler);
unblock_task(blocking_ctx);
\end{lstlisting}%
}
\caption{Pause and resume pattern to handle a synchronous operation}
\label{code:pause-resume-pattern}
\end{figure*}

The general usage pattern consists in replacing blocking operations by either asynchronous or non-blocking equivalents, and letting the runtime perform the actual blocking.
The runtime can then schedule other computations during the blocking period.
This usage scheme is shown in Figure~\ref{code:pause-pattern}.

Asynchronous operations that support callbacks can use the callback function to unblock the task.
If the operation does not support callbacks, then another thread has
to (1) periodically test for its completion and (2) unblock the tasks when it finishes.
Figure~\ref{code:resume-pattern} shows the pattern that the body of the main loop of such a thread would contain.
Notice that the information that associates an asynchronous operation with a blocking context must be made visible to the thread that will unblock it.

\subsection{Polling}
\label{sec:polling}

The detection of finished operations can be either blocking or non-blocking.
To simplify the non-blocking case, we propose an additional API that avoids the need for an additional thread.
Instead, the runtime can address those actions at regular intervals or on a best-effort basis.

To make this part generic, the API provides a periodic callback mechanism.
The callback should check for the completion of the asynchronous operation and perform the calls to unblock the associated task.
The prototype to register the callback is the following:

\begin{lstlisting}
    void register_polling_service(char const *service_name,
        polling_service_t service_function, void *service_data);
\end{lstlisting}

It receives a string parameter that is a description for debugging purposes, the callback function and an opaque pointer to data to pass to the callback.
The prototype of the callback is the following:

\begin{lstlisting}
    typedef int (*polling_service_t)(void *service_data);
\end{lstlisting} 

It receives as a parameter the opaque pointer, and returns a boolean value that indicates whether its purpose has been attained:
if true, the callback is automatically unregistered; otherwise the runtime will continue to call it.
Throughout the rest of this text we will refer to \emph{callback} as the pair composed by the callback function and the opaque data passed to the registration function.

During finalization, the following function can be used to unregister a callback.
It receives the same parameters as the registration function and returns once the callback has been disabled:

\begin{lstlisting}
    void unregister_polling_service(char const *service_name,
        polling_service_t service_function, void *service_data);
\end{lstlisting}

\subsection{External Events}
\label{sec:external_events_api}
We propose a generic API to bind the release of task dependencies to the completion of external events.
The aim of this API is to support the efficient execution of non-blocking\slash asynchronous operations within tasks, in which
the release of dependencies of the calling tasks are bound to the completion of those operations.
It is composed of three functions. The first has the following prototype in C:
\begin{lstlisting}
	void *get_current_event_counter();
\end{lstlisting}

This function returns an opaque pointer to runtime-specific data, which we call an \emph{event counter} throughout the rest of this text.
The binding of new external events is done through the following function:
\begin{lstlisting}
	void increase_current_task_event_counter(void *event_counter,
		unsigned int increment);
\end{lstlisting}

This function atomically increases the number of pending external events of the calling task.
The first parameter of the function must be the event counter of the invoking task, while the second parameter is the number
of external events to be bound.
The presence of pending events in a task prevents the release of its dependencies, even if the task has finished its execution.
Note that only the task itself can bind its external events.

\begin{lstlisting}
	void decrease_task_event_counter(void *event_counter,
		unsigned int decrement);
\end{lstlisting}

Then, the function shown above atomically decreases the number of pending external events of a given task.
The first parameter is the event counter of the target task, while the second parameter is the number
of completed external events to be decreased.
Once the number of external events of the task becomes zero and the task finishes its execution,
the task can release its dependencies.

Note that, all the external events of a task can complete before it actually finishes its execution.
In this case, the task will release its dependencies as soon as it finishes its execution.
Otherwise, the last call that makes the counter become zero, will trigger the release of the dependencies.

\begin{figure*}
 \centering
 \subfloat[Code that binds the asynchronous operation as an external event]{%
 \label{code:register-pattern}%
\begin{lstlisting}[numbers=left,numberstyle=\tiny,numbersep=5pt]
async_handler = start_async_op(...);
void *event_cnt = get_current_event_counter();
increase_current_task_event_counter(event_cnt, 1);
associate(async_handler, event_cnt);
// Do other things and finish the execution ...
\end{lstlisting}%
}
\hfil
\subfloat[Body of the code that handles the fulfillment of an external event]{%
 \label{code:unregister-pattern}%
\begin{lstlisting}[numbers=left,numberstyle=\tiny,numbersep=5pt]
async_handler = wait_until_one_async_op_finishes();
void *event_cnt = get_assigned_event_counter(async_handler);
decrease_task_event_counter(event_cnt, 1);
\end{lstlisting}%
}
\caption{External event's binding and fulfillment pattern to handle an asynchronous/non-blocking operation}
\label{code:register-unregister-pattern}
\end{figure*}

The general usage pattern consists in binding an external event to the task that performs an asynchronous or non-blocking operation, and to prevent the release of its dependencies until the operation has completed.
The task can finish its execution and free its stack, both without waiting for the completion of the asynchronous operation.
Figure~\ref{code:register-pattern} shows this usage scheme.

Asynchronous operations that support callbacks can use the callback function to fulfill the external event of the corresponding task.
If the operation does not support callbacks, then another thread has
to (1) periodically test for its completion and (2) fulfill the external event of the task once the operation has finished.
Note that the fulfillment of a task's external event could trigger the release of its task dependencies.
Figure~\ref{code:unregister-pattern} shows the pattern that the body of the main loop of such a thread would contain.
Also, notice that the information that associates an asynchronous operation with an event counter must be made visible to the thread that will fulfill the event (i.e., by decreasing the counter).

In addition, this external events API can be combined with the polling services API (Section~\ref{sec:polling}) similarly as the task pause/resume API does~\cite{interopmpi}.

\subsection{Blocking and Unblocking in Nanos6}

The blocking call in Nanos6 forces a scheduling point in the task.
At this point, the task will not be able to resume until it is sent back to the scheduler.
If there are ready tasks, the scheduler will assign one to the core.
Otherwise, the core will become idle.

The unblocking call sends the task back to the scheduler.
During this process, the scheduler may choose to wake up an idle core and assign the task to it.
In that case, the runtime resumes the execution on that core.
Otherwise, the task will eventually resume when there is a core available for it.

\subsection{Polling in Nanos6}

Nanos6 invokes the polling callbacks both at periodic intervals and opportunistically. 
The runtime has a thread dedicated to management operations, which processes the list of callbacks every 1 ms.
Performing calls at regular intervals allows it to support implementations that require them to guarantee progress.

In addition, worker threads serve the list of callbacks before letting their core become idle.
The implementation allows several threads to process the list concurrently.
However, at this time we assume that callbacks may not support concurrent execution.

The current polling API and its implementation in Nanos6 are at an early stage.
In the future we may add options related to callback concurrency and quality of service requirements.

\subsection{External Events in Nanos6}
Nanos6 tasks incorporate an atomic counter to determine when their dependencies can be released.
The runtime system initializes the counter to 1 to prevent the release of dependencies while the task is running.
The task itself is the only one allowed to increase its counter by calling to the {\it increase} function.

Finally, the runtime system automatically releases the dependencies of a task when the counter becomes zero, which can be either when the task finishes its execution (i.e., by decreasing the counter by 1), or later when somebody calls the {\it decrease} function.

\subsection{Genericity}

While we focus on OpenMP tasks and MPI, these APIs can also be applied to other task-based programming models and even other OpenMP work-sharing constructs.
For instance, an OpenMP runtime could execute more parallel loop iterations while others are blocked on MPI calls.
The APIs also support other types of operations with blocking and asynchronous variants, e.g., file accesses.

%% file: tex/mpi.tex
\section{MPI Progress}
\label{sec:mpi}

Calling blocking MPI primitives from within tasks can easily produce deadlocks when there are no
measures in place to constrain their execution order (e.g., through task dependencies).
For instance, consider a single thread in a single process executing a
task-based runtime with two tasks---one calls a blocking
synchronous-mode send and the other calls a blocking receive
(these calls match and there are no other calls that can match).
In MPI, as it is today, this must certainly deadlock and is therefore erroneous by definition.
Whichever order the blocking send and blocking receive are executed in,
either the {\sf MPI\_Ssend} will block until the {\sf MPI\_Recv} is issued, or vice versa.
The only execution thread available cannot proceed to the second call without completing the first.

However, with the task pause/resume API, the first blocking function call could pause the calling task,
enter the task-based runtime, schedule the other task, issue the second MPI blocking function call,
complete it because now both send and receive have been posted,
and then return and resume the first task, which can complete its MPI function.

%% file: tex/tampi.tex
\section{Task-Aware MPI Library}
\label{sec:interop_lib}

The Task-Aware MPI (TAMPI) library implements mechanisms for improving
the interoperability between task-based programming models and both blocking and non-blocking MPI operations.
TAMPI works on top of any MPI implementation that supports the {\sf MPI\_\-THREAD\_\-MULTIPLE} threading level.
Throughout the rest of this text, we call the blocking mode to the mechanism targeting the blocking MPI primitives,
and the non-blocking mode to the mechanism targeting the non-blocking MPI primitives.

\subsection{Blocking Mechanism}
\label{sec:blocking_mode}
The blocking mode prevents the underlying hardware thread, which executes a task that calls blocking MPI operations, from blocking
inside the MPI library, and allows it to execute other tasks instead, while the operation does not complete.
This mechanism leverages the standard MPI interception techniques that enable transparent interception of MPI calls
performed by an application, as well as, both the task pause/resume and polling services APIs introduced in Section~\ref{sec:apis}.

Figure~\ref{code:mpi_receive} shows the code that is executed when an application performs an {\sf MPI\_Recv} call from inside a task.
The first operation performed at line 3 is to check if the TAMPI's interoperability is enabled.
If this is not the case, the original blocking {\sf MPI\_Recv} operation is executed (line 15) using the PMPI interface.
Otherwise, the blocking call is transformed into its non-blocking counterpart, in this case an
{\sf MPI\_Irecv} (line 5). The code then checks if the operation has completed immediately.
In such case, it returns without blocking the task.
Otherwise, it creates a ticket object with the information about the ongoing MPI operation and the current
task (line 9).
The ticket is next registered inside the library and the task is paused (line 11).
MPI asynchronous operations do not feature a callback to resume the task once the operation is completed.
To handle this, the library defines a polling service callback (line 18), which the runtime system calls periodically to check if any MPI operation has completed (line 21).
When an MPI operation completes, the task waiting for that MPI operation is sent back to the runtime scheduler (line 24)  so that it can be resumed.
All other blocking MPI primitives, including collective operations, are intercepted and managed
similarly.

\setlength{\abovecaptionskip}{8pt}
\setlength{\belowcaptionskip}{-4pt}

\begin{figure}[t]
\centering
\begin{lstlisting}[numbers=left,numberstyle=\tiny,numbersep=5pt]
int MPI_Recv(void *buf, ..., MPI_Status *status) {
	int err, completed = 0;
	if (Interop::isEnabled()) {
		MPI_Request request;
		err = MPI_Irecv(buf, ..., &request);
		MPI_Test(&request, &completed, status);
		if (!completed) {
			Ticket ticket(&request, status);
			ticket._waiter = get_current_blocking_context();
			_pendingTickets.add(ticket);
			block_current_task(ticket._waiter);
		}
		return err;
	}
	return PMPI_Recv(buf, ..., status);
}

void Interop::poll() {
	for (Ticket &ticket : _pendingTickets) {
		int completed = 0;
		MPI_Test(ticket._request, &completed, ticket._status);
		if (completed) {
			_pendingTickets.remove(ticket);
			unblock_task(ticket._waiter);
		}
	}
}
\end{lstlisting}
\caption{Implementation of the {\sf MPI\_Recv} function and the polling function executed periodically by the runtime system in TAMPI}
\label{code:mpi_receive}
\end{figure}

\subsection{Non-Blocking Mechanism}
The non-blocking mode of TAMPI improves the blocking one by avoiding the pause of communication tasks.
The cost of blocking/unblocking tasks is sometimes not negligible.
Typically, the pause operation of a task produces a context switch, requires the runtime system to keep alive the stack of the paused task (e.g., 8 MB of stack size) and increases the scheduling overhead.
These overheads are significant in applications with many communication tasks or in which the communication messages are small.

Therefore, we propose two functions called {\sf TAMPI\_Iwait} and {\sf TAMPI\_Iwaitall}, which have the same parameters as the original
{\sf MPI\_Wait} and {\sf MPI\_Waitall} functions, respectively.
These two are non-blocking asynchronous functions that bind the release of the calling task's dependencies with the completion of the MPI
requests passed as parameters.
Once the calling task finishes its execution and all MPI operations registered with {\sf TAMPI\_Iwait}/{\sf TAMPI\_Iwaitall} complete,
the task's dependencies are automatically released.
For that, the library leverages both the task external events and polling services APIs detailed in Section~\ref{sec:apis}.

Since these new functions are asynchronous, we do not guarantee that after returning from one of these function calls the bound requests are already completed.
Data buffers involved in the communication will become available only when the task fully completes (i.e., after the release of its dependencies).
In addition, this non-blocking mode and the blocking mode presented previously are compatible so that they can coexist in the same application.

Figure~\ref{code:tampi_iwait} shows the code that is executed when the application calls {\sf TAMPI\_Iwait} inside a task.
As in the blocking mode, the first operation performed at line 3 is to check if the interoperability mechanism is enabled.
If this is not the case, the original blocking {\sf MPI\_Wait} function is executed (line~13).
Otherwise, the code checks if the MPI request passed as a parameter is immediately completed by calling to {\sf MPI\_Test} (line 4).
In such case, the function returns without increasing the external event counter of the calling task, since the MPI request has already been completed.
Otherwise, a ticket object is allocated and filled with the information about the ongoing MPI operation and the
event counter of the current task (line 7).
Note that unlike in the blocking mode, the ticket object cannot reside in the stack of the task, since the function will return immediately.

Then, a new pending event is bound to the current task by increasing its event counter (line 8), avoiding the release of its dependencies until it finishes its execution and all its pending events complete.
Finally, the ticket is registered inside the library (line 9), and the function returns immediately.

The polling function is similar to the one presented in the blocking mode.
Once a ticket completes, the event counter of the task that owns that ticket is decreased (line 22).
If this was the last pending event and the task already finished its execution, its dependencies are automatically released by the runtime system.

\begin{figure}
\centering
\begin{lstlisting}[numbers=left,numberstyle=\tiny,numbersep=5pt]
int TAMPI_Iwait(MPI_Request *request, MPI_Status *status) {
	int err, completed = 0;
	if (Interop::isEnabled()) {
		err = MPI_Test(request, &completed, status);
		if (!completed) {
			Ticket *ticket = new Ticket(request, status);
			ticket->_cnt = get_current_event_counter();
			increase_current_task_event_counter(ticket->_cnt, 1);
			_pendingTickets.add(ticket);
		}
		return err;
	}
	return PMPI_Wait(request, status);
}

void Interop::poll() {
	for (Ticket &ticket : _pendingTickets) {
		int completed = 0;
		MPI_Test(ticket._request, &completed, ticket._status);
		if (completed) {
			_pendingTickets.remove(ticket);
			decrease_task_event_counter(ticket._cnt, 1);
			delete ticket;
		}
	}
}
\end{lstlisting}
\caption{Implementation of the {\sf TAMPI\_Iwait} function and the polling function executed periodically by the runtime system in TAMPI}
\label{code:tampi_iwait}
\end{figure}

Figure~\ref{code:tampi_iwaitall_example} shows an example on how to use the new non-blocking asynchronous {\sf TAMPI\_Iwaitall} function in conjunction with
other asynchronous MPI primitives within a task.
The first task declares the proper dependencies on the data buffers, which are used to receive/send data by calling the non-blocking MPI send/receive
operations.
Then, the task binds the produced MPI requests to the release of its dependencies by calling {\sf TAMPI\_Iwaitall}.
As previously stated, the running task cannot consume/reuse the data buffers, since there is no guarantee that the requests are completed after {\sf TAMPI\_Iwaitall} returns.

The second task declares an input dependency on the integer received by the communication task and prints it.
This last task will become ready once the communication task finishes its execution and both MPI requests complete.

\begin{figure}
\centering
\begin{lstlisting}[numbers=left,numberstyle=\tiny,numbersep=5pt]
#pragma oss task out(buf[0]) in(buf[1]) out(statuses[0;2])
{
	MPI_Request reqs[2];
	MPI_Irecv(&buf[0], 1, MPI_INT, r, t, MPI_COMM_WORLD, &reqs[0]);
	MPI_Isend(&buf[1], 1, MPI_INT, r, t, MPI_COMM_WORLD, &reqs[1]);
	TAMPI_Iwaitall(2, reqs, statuses);
	// Data buffer buf[0:1] cannot be consumed/reused here!
}

#pragma oss task in(buf[0]) in(statuses[0])
{
	check_status(statuses[0]);
	printf("Received integer: %d\n", buf[0]);
}
\end{lstlisting}
\caption{Example using the new {\sf TAMPI\_Iwaitall} function in conjunction with other asynchronous MPI primitives}
\label{code:tampi_iwaitall_example}
\end{figure}

\subsection{New Thread Support Level}
\label{sec:threading-level}
We propose that MPI should define a new thread support level,
which each MPI library can choose to support or not during initialization of MPI.
The new thread support level could be called {\sf MPI\_TASK\_MULTIPLE} and 
its constant value would be monotonically greater than the {\sf MPI\_THREAD\_MULTIPLE} constant.
In this way, applications can request support for the interoperability functionality via the {\sf MPI\_Init\_thread} call and check whether the underlying MPI library provides it.
More information on how the new threading level should be defined in the MPI Standard is included in~\cite{interopmpi}.

In the case of the TAMPI library, both blocking and non-blocking interoperability mechanisms are enabled when initializing MPI with this new threading level.
Figure~\ref{code:mpi_task_multiple} shows an example of how a hybrid MPI+OmpSs code
may use this new thread support level to write portable applications.
First, the application checks if the {\sf MPI\_TASK\_MULTIPLE} threading level is supported by the underlying MPI library.
If this is the case, it defines a sentinel variable pointing to NULL, which will be ignored by the OmpSs-2 dependency system.
Otherwise, it sets the sentinel variable to a non-NULL value, so that communication tasks will be serialized.
This is shown in lines 11--12, where communication tasks are
created with a regular dependency over the block these will work
on, as well as an artificial {\it inout} dependency on the memory position pointed by the sentinel variable to serialize the execution of these tasks and avoid deadlocks.
Nevertheless, if the sentinel points to NULL, communication tasks do not enforce the artificial dependency so that they can run in parallel.

\begin{figure}
\centering
\begin{lstlisting}[numbers=left,numberstyle=\tiny,numbersep=5pt]
int *sentinel; // Sentinel used to serialize communication tasks

int main(int argc, char * argv[]) {
	int provided;
	MPI_Init_thread(&argc, &argv, MPI_TASK_MULTIPLE, &provided);
	if (provided == MPI_TASK_MULTIPLE) sentinel = NULL;
	else sentinel = (int *) 1;

	for (int i=0; i<NT; i++) {
		// Dependency enforced only if *sentinel != NULL
		#pragma oss task inout(tile[i]) inout(*sentinel)
		communication_task(tile[i]);
	}
}
\end{lstlisting}
\caption{Portable initialization using {\sf MPI\_TASK\_MULTIPLE}}
\label{code:mpi_task_multiple}
\end{figure}

%% file: tex/evaluation.tex
\section{Evaluation}
\label{sec:evaluation}

In this section we provide an in-depth evaluation of the programmability and performance of our proposal to improve MPI and OpenMP interoperability.
We analyze our results based on two benchmarks: an iterative Gauss--Seidel method and a mock-up of a meteorological forecasting application. 
We have used up to 64 compute nodes of the MareNostrum 4 supercomputer to run the experimental validation.
Each compute node is equipped with 2 sockets of Intel Xeon Platinum 8160 CPUs, with 24 cores each, totaling 48 cores per node, and 96~GB of main memory.
The interconnection network is based on 100~Gbit/s Intel Omni-Path HFI technology. 
We have used the latest stable release of MPICH (3.2.1) and OmpSs-2 (18.06).

\subsection{Gauss--Seidel}

In this section we use the iterative Gauss--Seidel method~\cite{Greenbaum:1997:IMS:264985} to solve the Heat equation \cite{Esposito2017}, which is a parabolic partial differential equation that describes the distribution of heat in a given region over time.
We have developed five versions of the Gauss--Seidel method for 2-D matrices. The following two are MPI-based:

\begin{figure}
\centering
	\includegraphics[width=.80\linewidth]{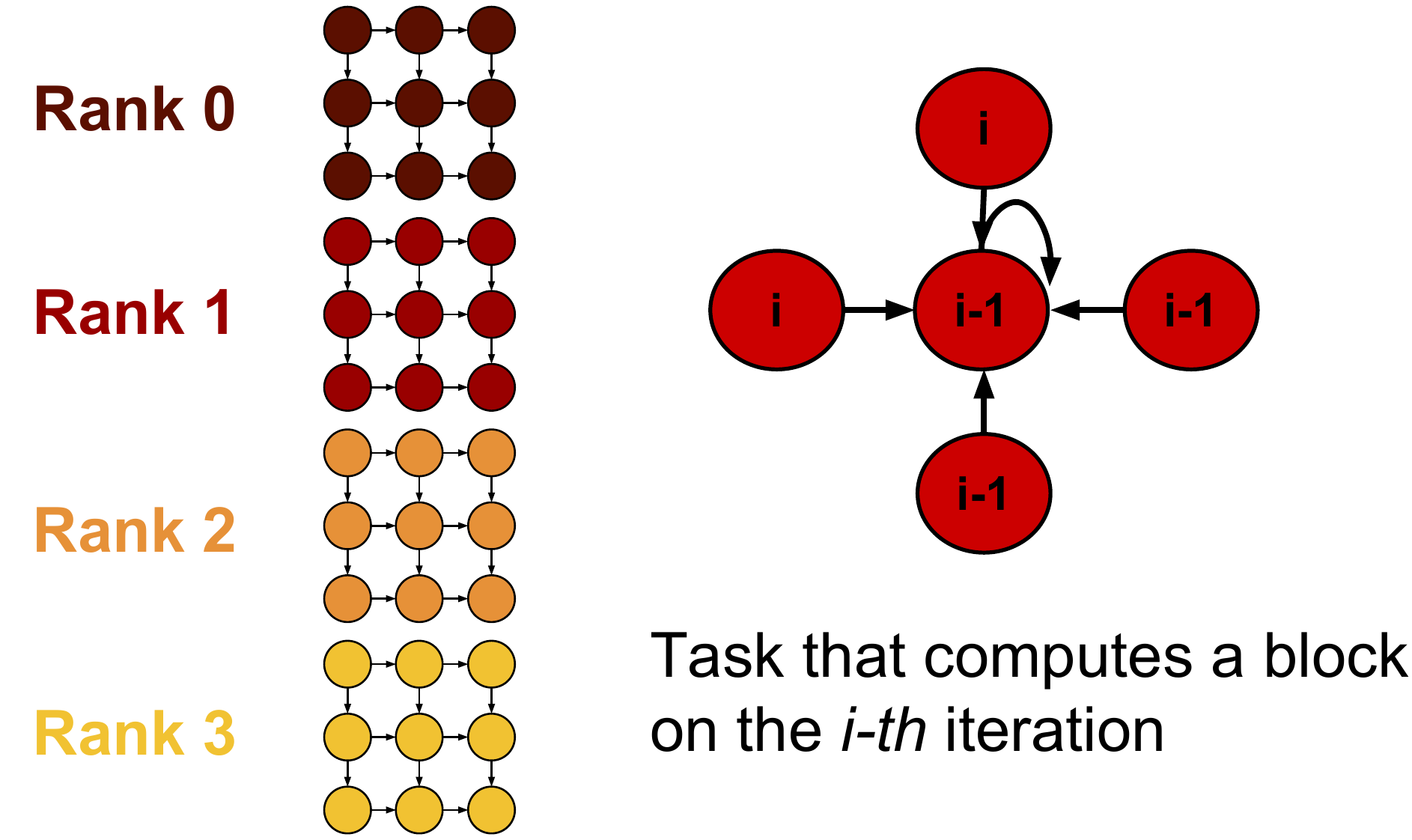}
\caption{2-D matrix of $3 \times 12$ blocks split in four ranks. On the hybrid versions, for each iteration a task is created to update each block using values of both current (top and left blocks) and previous (current, right and bottom blocks) iterations}
\label{fig:gs_diag}
\end{figure}

\begin{itemize}[leftmargin=*]
\item {\it Pure MPI}: This version is a straightforward implementation of the algorithm using synchronous MPI primitives to exchange boundaries among neighbouring ranks. The computation phase of the algorithm is sequential.
The 2-D matrix is distributed across ranks assigning a consecutive set of rows to each one (a single block per rank). Boundary exchanges correspond to whole rows.

\item {\it N-Buffer MPI}: This version is significantly more elaborate than {\it Pure MPI}.
In this case the rows of each rank are horizontally divided by blocks; hence a distinct boundary exchange is performed for each block.
This version starts to exchange block boundaries as soon as possible using asynchronous MPI primitives.
For instance, a rank starts to send ({\sf MPI\_Isend}) its last row of
a block once it has been computed, but it also starts to receive ({\sf MPI\_Irecv}) the lower boundary for the next iteration.
Before starting the computation of a block, it waits ({\sf MPI\_Wait}) for the completion of all pending MPI requests related to the block.
Thus, the computation is partially overlapped by boundary exchanges.
\end{itemize}

The rest are hybrid MPI+OmpSs versions which divide the matrix into squared blocks and these are distributed across MPI ranks.
The left-hand side of Figure~\ref{fig:gs_diag} shows how a domain of
$3 \times 12$ blocks would be split across four MPI ranks.
These hybrid versions are:

\begin{itemize}[leftmargin=*]
\item {\it Fork-Join}: This is a hybrid version with a sequential communication phase and a parallel computation phase.
The communication phase uses synchronous primitives to exchange boundaries among neighbours as in {\it Pure MPI}.
On the computation phase, a task is created to update each block using the top and left blocks of the current iteration, and the current, left and bottom blocks of the previous iteration, as shown in Figure~\ref{fig:gs_diag}.
Tasks use fine-grained dependencies to exploit the spatial wave-front parallelism.
However, there is a global synchronization point after each computation phase that prevents this version from exploiting parallelism across iterations (temporal wave-front).
	
\item {\it Sentinel}: A hybrid version where both communication and computation are implemented using tasks.
The communication phase uses tasks to execute the synchronous MPI primitives that exchange boundary blocks among neighbours.
These communication tasks are serialized by a sentinel dependency to avoid deadlocks (as discussed in Section~\ref{sec:mpi}).
This version avoids the global synchronization ({\it taskwait}) by leveraging fine-grained dependencies between computation and communication tasks.

\item {\it Interop(blk)}: This version uses the {\sf MPI\_TASK\_MULTIPLE} multithreading level proposed in this paper to avoid the serialization of communication tasks.
This is the only difference with the {\it Sentinel} version.
It has been evaluated with the blocking mode of the TAMPI library (Section~\ref{sec:interop_lib}), which is the one that leverages the task pause/resume API.
\end{itemize}

In {\it N-Buffer MPI}, each block has {\it total\_rows/num\_ranks} rows and 1K columns.
In the hybrid versions, each compute task processes a block of 1K
$\times$ 1K elements. This is the smallest block size required to attain peak performance.

\begin{figure}
\centering
\includegraphics[height=7.9cm, width=1\linewidth]{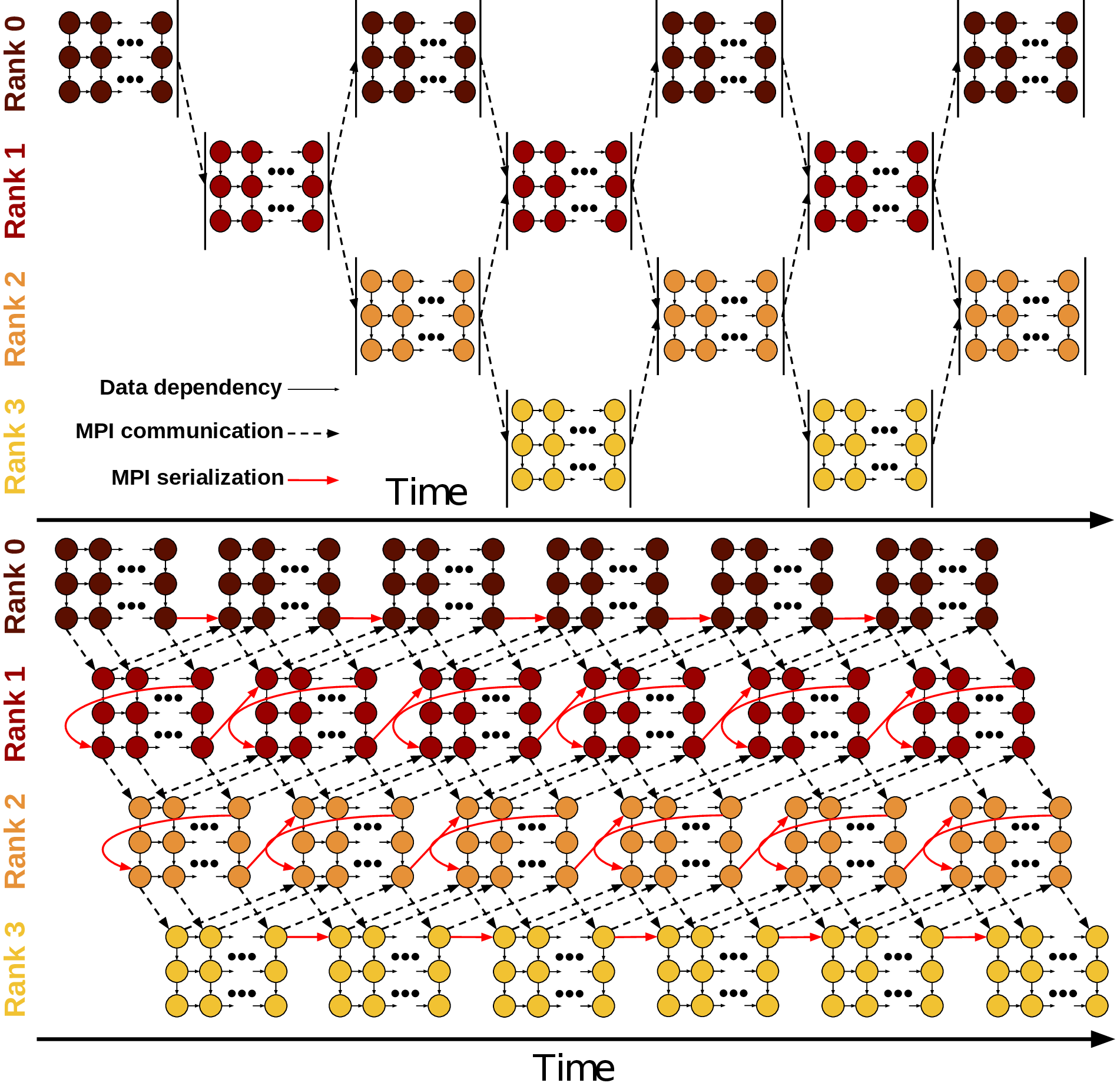}
\caption{Above: dependency graph for {\it Pure MPI} and {\it Fork-Join}. Below: dependency graph for {\it N-Buffer MPI}, {\it Sentinel} (with red dependencies) and {\it Interop} (no red dependencies)}
\label{fig:gs_flow}
\end{figure}

Figure~\ref{fig:gs_flow} compares the dependency graph of {\it Pure MPI} and {\it Fork-Join} (above) and {\it N-Buffer MPI}, {\it Sentinel} and {\it Interop} (below).
For the sake of clarity, both graphs have been simplified by showing up to the first six iterations, fusing the explicit communication tasks with the tasks that compute boundary blocks and also other redundant dependencies such as anti-dependencies.
In the {\it Pure MPI} and {\it Fork-Join} versions, the execution of each iteration inside an MPI rank depends on the completion of the previous iteration of its neighbor MPI ranks, which results in a strong serialization effect that affects the execution of the whole program.

In the {\it N-Buffer MPI} version, the strong serialization effect can be avoided by exchanging block boundaries as soon as possible, performing calls to the corresponding asynchronous MPI primitives right after processing each block.
The {\it Sentinel} version also exchanges block boundaries at the earliest, using tasks with fine-grained dependencies to execute MPI primitives.
However, since this version uses synchronous primitives, it still has to serialize communication tasks to avoid deadlocks. This introduces the red dependencies, shown in the lower part of Figure~\ref{fig:gs_flow}, that also reduce significantly the parallelism within and across iterations.

Finally, the {\it Interop(blk)} version that uses the new {\sf MPI\_\-TASK\_\-MULTIPLE} threading level removes the red dependencies and thus can fully exploit both spatial and temporal wave-front parallelism.
Moreover, in this version, tasks blocked on MPI calls never block the underlying hardware thread, so resource undersubscription is also avoided.
In summary, the proposed threading level allows the programmer to parallelize applications in a more natural way, without requiring artificial dependencies that hinder the available parallelism.

\begin{figure}
\centering
\includegraphics[width=1\linewidth]{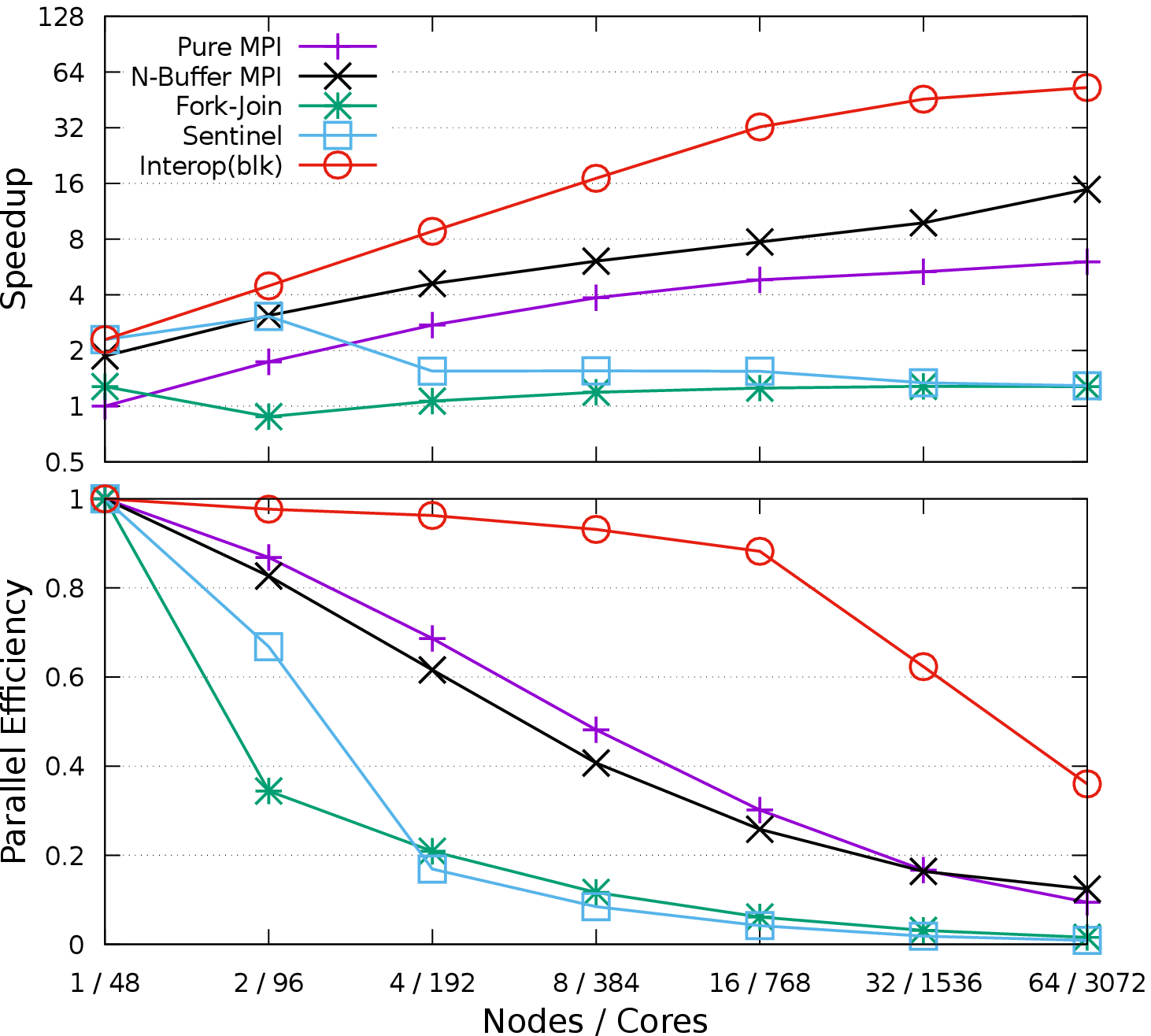}
\caption{Speedup and parallel efficiency of the Gauss--Seidel strong scaling (64K x 64K total elements; 1,000 iterations)}
\label{fig:gs_ss}
\end{figure}

{\it Pure MPI} and {\it N-Buffer MPI} experiments have been performed using 48 MPI ranks per node. Hybrid versions have used 1 rank per node and 48 OmpSs threads per rank. The upper part of Figure~\ref{fig:gs_ss} shows a strong-scaling study of the five versions using the performance of {\it Pure MPI} running on one node as a baseline.
On a single node, all hybrid versions experience higher performance than {\it Pure MPI}.
When the hybrid versions run on a single node (one rank), the MPI primitives are completely avoided.
Thus, the rigid serialization effect introduced by MPI is fully removed and these versions can fully exploit the spatial and temporal wave-front parallelism.
It is worth noting that the {\it Fork-Join} version is significantly slower than the other task-based versions due to the global synchronization point after each iteration that prevents the exploitation of the temporal wave-front.
As we increase the number of nodes, the performance of the {\it Pure MPI} version also increases, but the scalability is clearly sub-optimal. On the other hand, both {\it Fork-join} and {\it Sentinel} stop scaling at two and four nodes, respectively.
Note that these versions are the only that can be easily implemented with current OpenMP and MPI standards. 

The {\it N-Buffer MPI} version outperforms all previous versions, since it avoids the strong serialization of iterations among ranks which is observed in {\it Pure MPI} and {\it Fork-Join}.
In addition, this version allows the overlapping of computation and communication phases.
However, the scalability is still sub-optimal and the algorithm is difficult to implement.

The {\it Interop(blk)} version experiences good scalability up to 32 nodes.
With 64 nodes the curve flattens because the problem size is too small to get sufficient parallelism to exploit 48 cores.

The lower part of Figure~\ref{fig:gs_ss} shows the parallel efficiency of all five versions.
In this case each version uses as a baseline its own performance on a single node.
From 1 to 16 nodes the efficiency of the {\it Interop(blk)} version is almost the same, but then it quickly decreases, since the problem size becomes too small to feed all the cores.
The parallel efficiency of {\it Pure MPI} and {\it N-Buffer MPI} steadily decreases from 1 to 0.1 at 64 nodes.
{\it Fork-Join} and {\it Sentinel} have a big drop of parallel efficiency at two and four nodes, respectively.
\begin{figure}
\centering
  \subfloat[Pure MPI\label{fig:trace_pure_mpi}]{%
       \centering\includegraphics{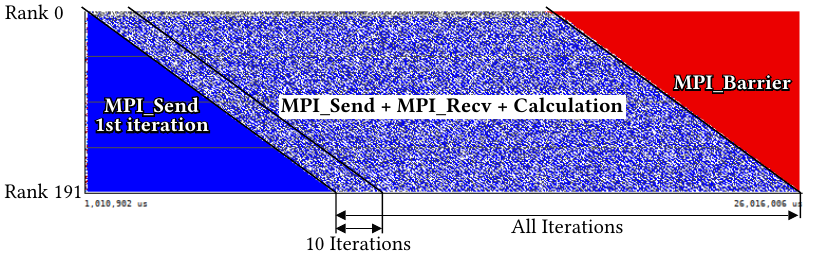}
  }
  \hfill
  \subfloat[N-Buffer MPI\label{fig:trace_nbuffer_mpi}]{%
       \centering\includegraphics{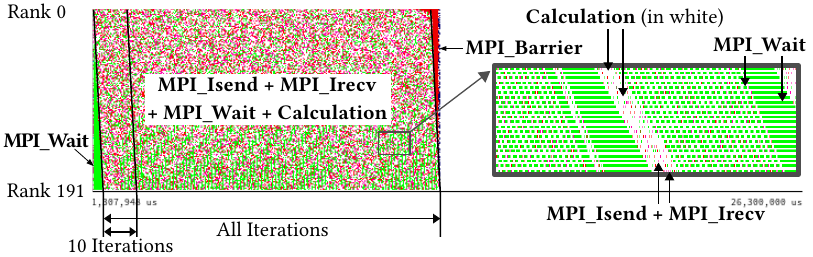}
  }
  \hfill
  \subfloat[Fork-join\label{fig:trace_fork_join}]{%
       \centering\includegraphics{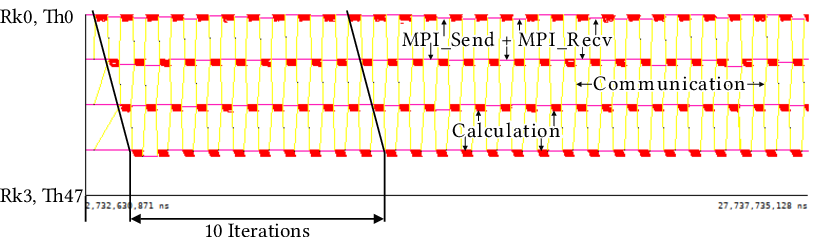}
  }
  \hfill
  \subfloat[Sentinel\label{fig:trace_sentinel}]{%
       \centering\includegraphics{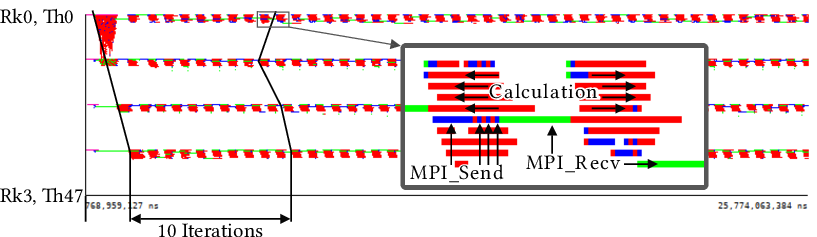}
  }
  \hfill
  \subfloat[Interop(blk)\label{fig:trace_interop}]{%
       \centering\includegraphics{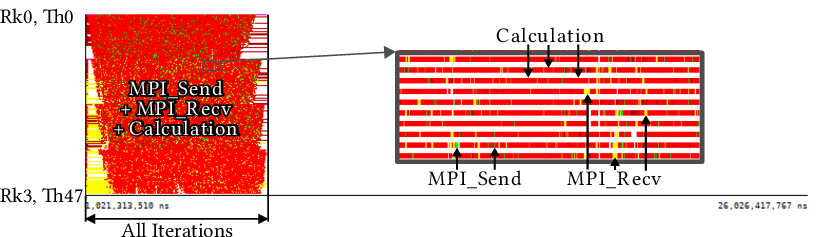}
  }
\caption{Execution traces with 4 nodes. The Y axis shows MPI ranks/OmpSs threads and the X axis is the time-line}
\label{fig:traces}
\end{figure}

Figure \ref{fig:traces} shows five traces of {\it Pure MPI}, {\it N-Buffer}, {\it Fork-Join}, {\it Sentinel} and {\it Interop(blk)}, respectively, running on four nodes (192 cores) with the same time-scale.
The traces show the time-line on the X axis and the MPI ranks/OmpSs threads on the Y axis.
In {\it Pure MPI} and {\it N-Buffer} there are 192 ranks; in the other versions there are four ranks---one rank per node---and each rank has 48 OmpSs threads.
On the three hybrid versions, the red lines correspond to the execution of the Gauss--Seidel tasks.

On the {\it Pure MPI} version~(Figure~\ref{fig:trace_pure_mpi}) the last rank (191) cannot start computing the first iteration until all the other ranks have completed the first iteration.
This introduces a big delay at the beginning that is also symmetrically reproduced at the end.

The same effect can be observed on the {\it Fork-Join}~(Figure~\ref{fig:trace_fork_join}) and {\it Sentinel}~(Figure~\ref{fig:trace_sentinel}) versions, but in this case there are only four ranks, so only four full iterations are required to have all the MPI ranks working.
In the {\it Fork-Join} version, the global synchronization at the end of each iteration produces a strong serialization effect among iterations (that is the same effect found on the {\it Pure MPI} version), so one iteration cannot start until the same iteration of the previous MPI rank has been fully completed.
Moreover, the global synchronization at the end of the computation phase also limits the available parallelism, so only 8 out of the 48 cores can work in parallel (running computation tasks).

The {\it Sentinel} version improves over the {\it Fork-Join} version because one MPI rank can start computing an iteration as soon as the previous rank has completed the computation of the first boundary block of the same iteration.
This allows to partially overlap the computation of the same iteration across MPI ranks.
However, the artificial dependencies introduced to serialize the communication tasks still hinder the available parallelism inside one iteration.
In this case, 8 cores can run computation tasks in parallel with another core running a communication task.
Although these hybrid versions take less time to complete a single iteration, {\it Pure MPI} pipelines iterations in a better way and ends up outperforming them in overall iteration throughput.

{\it N-Buffer MPI}~(Figure~\ref{fig:trace_nbuffer_mpi}) does not show the big delay at the first iteration seen in {\it Pure MPI} and {\it Fork-Join}.
This is because it exchanges boundaries as soon as possible, thus ranks can process different blocks from the same iteration concurrently.
In addition, it is more flexible than the previous ones due to the use of asynchronous MPI primitives.
The aforementioned reasons make this version outperform previous versions both in iteration latency and overall iteration throughput.
However, it does not reach {\it Interop(blk)}'s performance and it requires more development effort than them.

Finally, the {\it Interop(blk)}~(Figure~\ref{fig:trace_interop}) version avoids any global synchronization or serialization of communication tasks, so an iteration can be almost fully overlapped across the ranks.
Moreover, this version is the only that can exploit both spatial
wave-front and temporal wave-front parallelisms, benefiting from the 48 cores.

\begin{figure}
\centering
\includegraphics[width=1\linewidth]{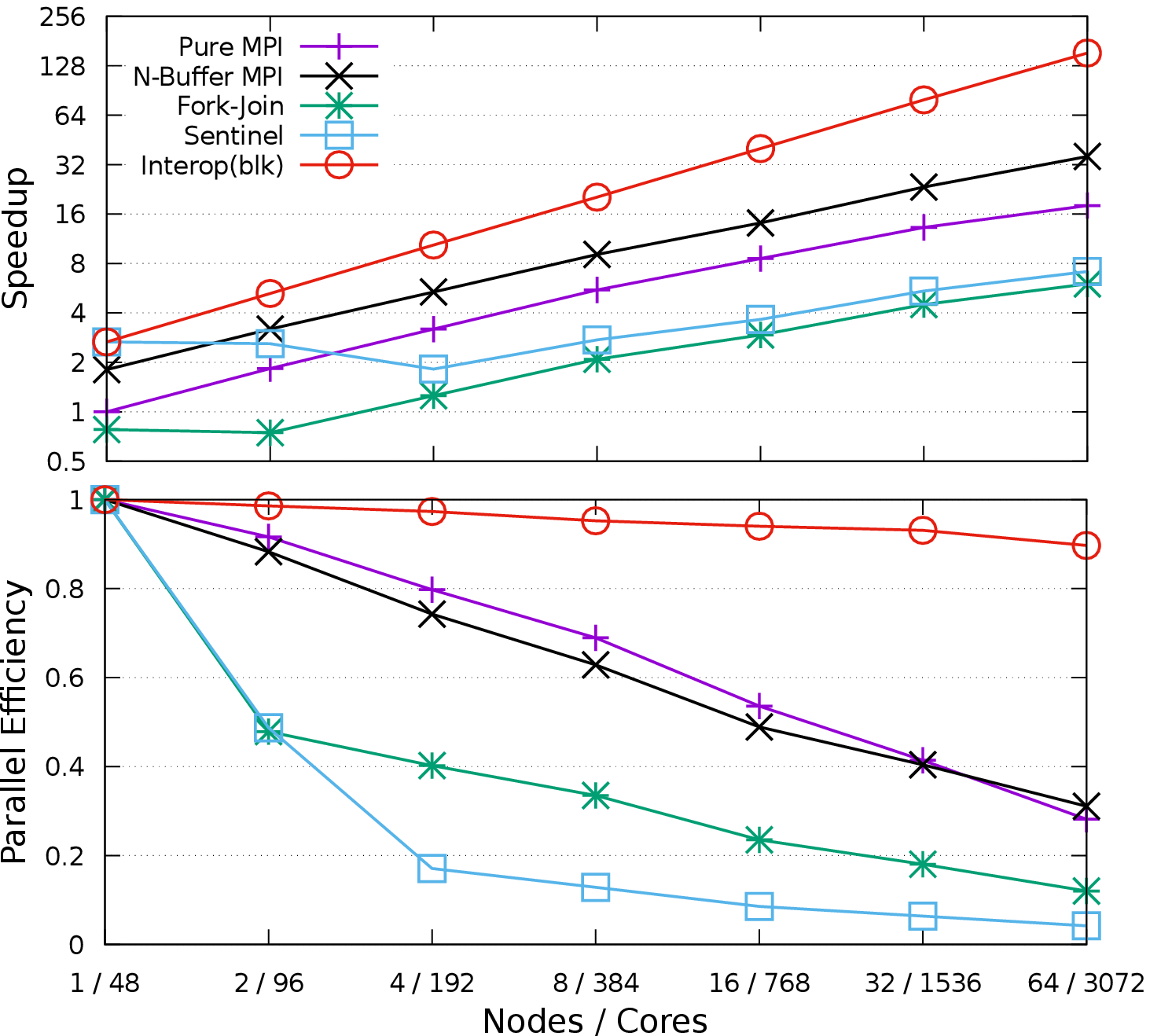}
\caption{Speedup and parallel efficiency of the Gauss--Seidel weak scaling (32K x 32K elements per node; 1,000 iterations)}
\label{fig:gs_ws}
\end{figure}

We have also performed a weak-scaling experiment.
The speed-up graph (upper part of Figure~\ref{fig:gs_ws}) uses the performance of the {\it Pure MPI} version on a single node as a baseline for all versions.
For the parallel efficiency graph (lower part of Figure~\ref{fig:gs_ws}), each version uses its own performance on one node as the baseline.
This experiment shows again the good scalability of the {\it Interop(blk)} version that scales linearly up to 64 nodes.
The parallel efficiency of {\it Pure MPI} and {\it N-Buffer MPI} steadily decreases from 1 to 0.3 at 64 nodes, while {\it Fork-Join}
and {\it Sentinel} feature a parallel efficiency of 0.4 and 0.2, respectively, with only four nodes.

In addition, we obtained similar results when performing these experiments in other systems, e.g., Cray ARCHER.

To finalize the performance analysis of the Gauss--Seidel benchmark, we have compared the performance of the blocking and non-blocking modes of the TAMPI library.
For this experiment we have used Intel MPI 2017 Update 4.
On the one hand, the {\it Interop(blk)} Gauss--Seidel version corresponds to the one that uses the blocking mode of the library, as in the previous experiments.
On the other hand, the {\it Interop(non-blk)} version corresponds to the one that leverages the non-blocking mode of the library.
This version is the same as {\it Interop(blk)} but replacing all blocking MPI calls by their non-blocking counterparts followed by a {\sf TAMPI\_Iwait} call passing the corresponding MPI request.
However, the dependency graph of {\it Interop(non-blk)} remains exactly the same as for the {\it Interop(blk)} version (Figure~\ref{fig:gs_flow}).

\begin{figure}
\centering
\includegraphics[width=1\linewidth]{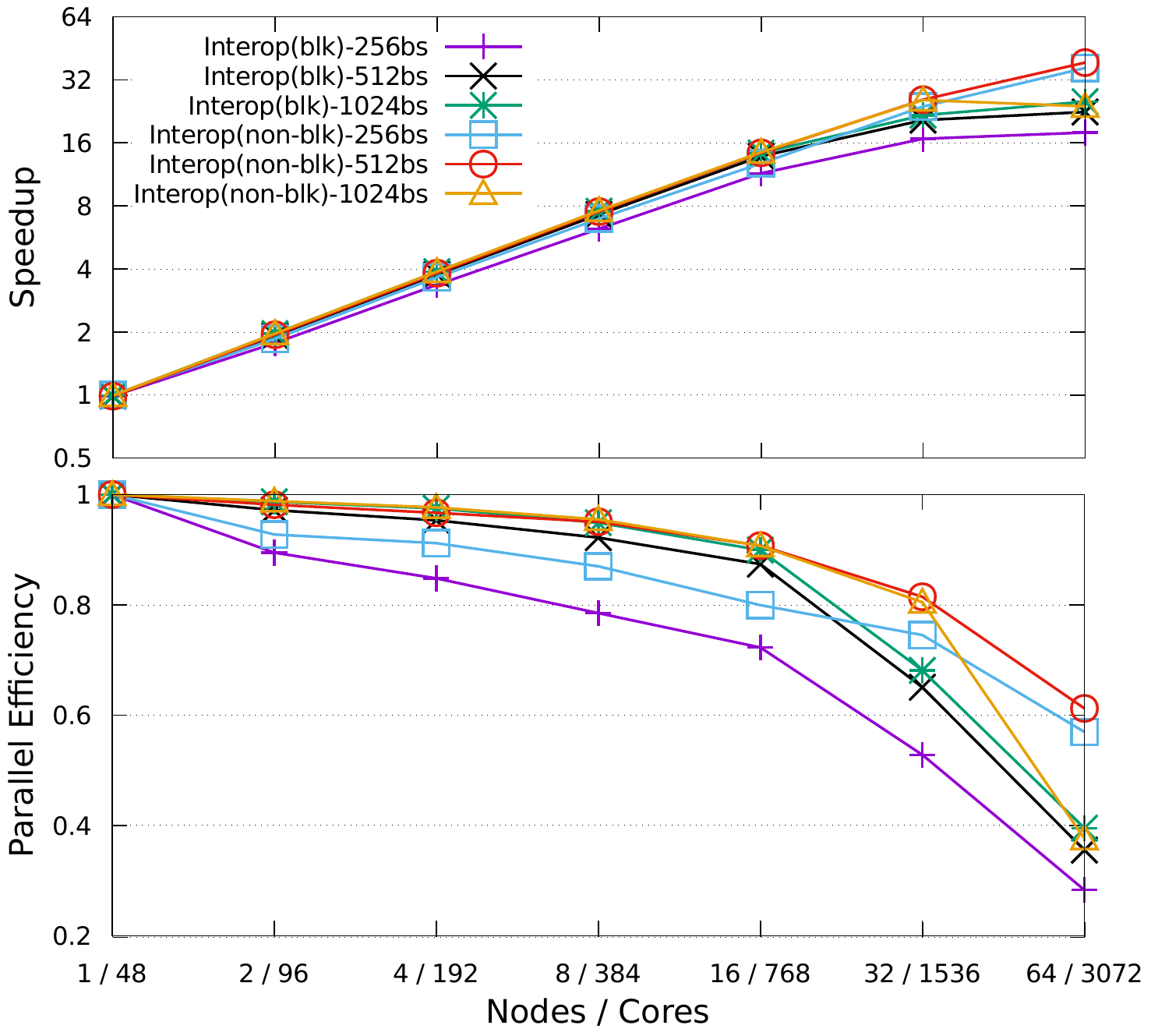}
\caption{Interop(blk) vs. Interop(non-blk): Speedup and parallel efficiency of the Gauss--Seidel strong scaling (64Kx64K total elements; 2,000 iterations)}
\label{fig:gs_ss_bnb}
\end{figure}

The top of Figure~\ref{fig:gs_ss_bnb} shows the speed-up graph of the strong scaling experiment.
Both versions have been executed with block sizes 256$\times$256, 512$\times$512 and 1K$\times$1K.
The {\it Interop(blk)-1024bs} is equivalent to the {\it Interop(blk)} version shown in Figure~\ref{fig:gs_ss}.
When using 64 nodes, this version stops scaling due to lack of parallelism.
We could get more parallelism by decreasing the block size, but as can be seen in Figure~\ref{fig:gs_ss_bnb}, {\it Interop(blk)} obtains worse performance when using smaller block sizes than 1K$\times$1K.

In contrast, the {\it Interop(non-blk)} version significantly improves the performance when using smaller block sizes.
Specifically, it obtains the best performance with block size 512$\times$512, which is closely followed by the execution with 256$\times$256.
Note that when decreasing the block size from 512$\times$512 to 256$\times$256, the number of created tasks is multiplied by 4, which makes the overhead of the task schedulers more noticeable.

The reasons behind the differences between {\it Interop(blk)} and {\it Interop(non-blk)} are that the latter does not require any context switch, it does not need to keep the stack of the task while the MPI operations are in flight, and it goes through the scheduler less times.
In summary, the non-blocking mode of the TAMPI library (i.e., the one that internally leverages the external events API) allows decreasing the block size so that we can scale up to more computing nodes.

The bottom of Figure~\ref{fig:gs_ss_bnb} shows the parallel efficiency of this strong scaling experiment.
This plot confirms that the {\it Interop(non-blk)} variants are the ones that take the most from the computational resources.

\begin{figure}
\centering
\includegraphics[width=1\linewidth]{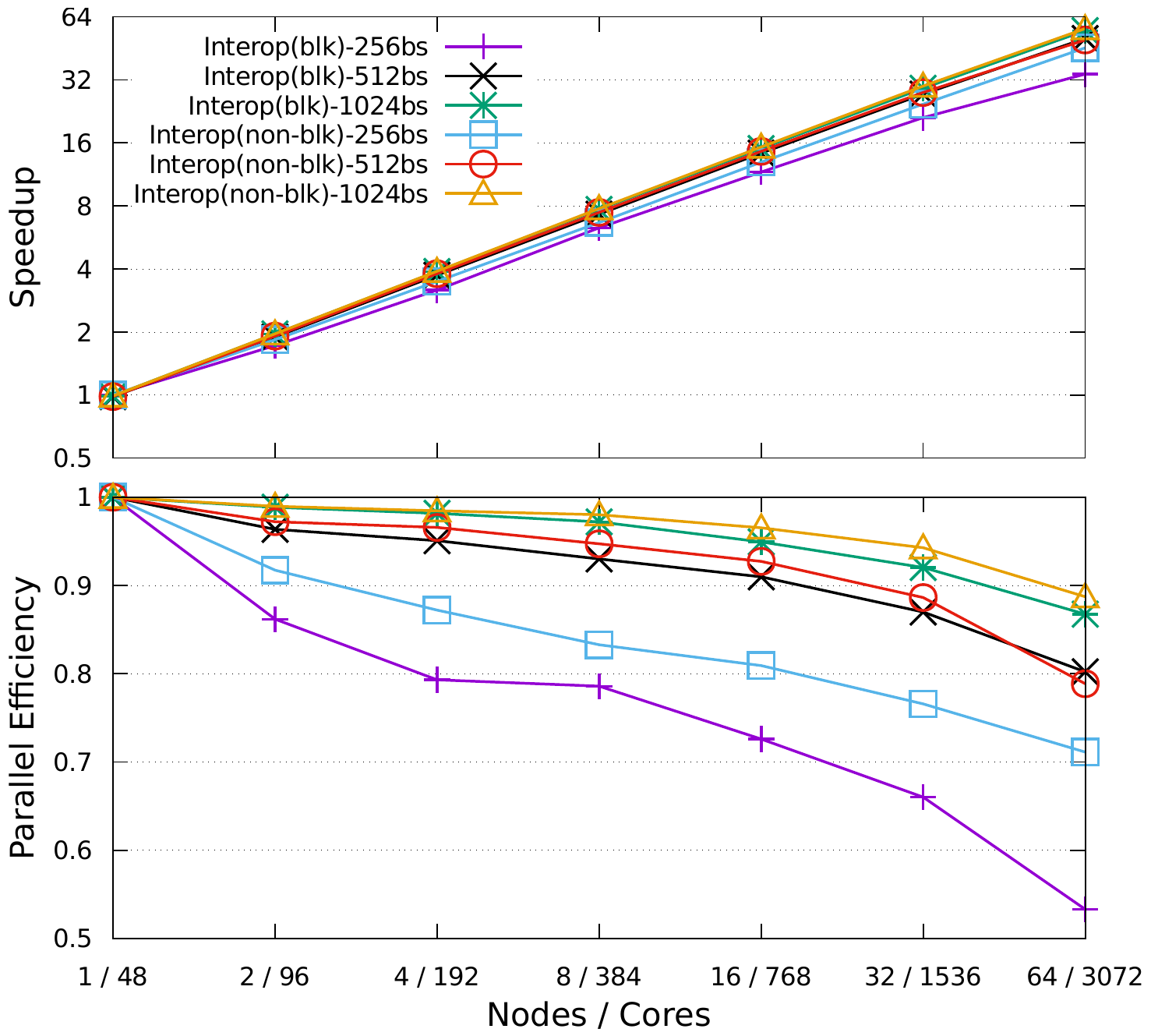}
\caption{Interop(blk) vs. Interop(non-blk): Speedup and parallel efficiency of the Gauss--Seidel weak scaling (32K x 32K elements per node; 1,000 iterations)}
\label{fig:gs_ws_bnb}
\end{figure}

Finally, we have executed a weak scaling experiment with both {\it Interop} versions and block sizes 256$\times$256, 512$\times$512 and 1K$\times$1K, which is shown in Figure~\ref{fig:gs_ws_bnb}.
In this case, there is no difference between {\it Interop(blk)} and {\it Interop(non-blk)} with block sizes 512$\times$512 and 1K$\times$1K, since the interoperability variants already scaled up almost linearly in the previous weak scaling experiments (i.e., Figure~\ref{fig:gs_ws}).
However, when decreasing the block size to 256$\times$256 (i.e., creating more tasks), the {\it Interop(blk)} version obtains significantly worse performance than {\it Interop(non-blk)}, since the latter avoids the overhead of blocking/unblocking communication tasks.

\subsection{IFSKer}
IFSKer is a mock-up application parallelized with MPI.
It mimics the communication and computational patterns of the meteorological forecasting model called Integrated Forecasting System (IFS).
IFS employs a spectral transform method which represents fields by using a set of coefficients of a basis function (e.g., a sine function).

The algorithmic structure consists of time-step cycles divided into two phases: grid-point physics computations and Fast Fourier transforms.
Data representation and distribution among MPI ranks is different in each stage.
Therefore, communication among ranks occurs during the transitions
among stages, where the data needs to be transposed and redistributed among the ranks.

The original implementation is based on MPI ({\it Pure MPI}), but we have implemented a new version ({\it Interop}) that uses tasks for both the computation and communication phases.
However, in this application the computation phase is very fine-grained, so it is not worth to fully parallelize it.
Hence, we only use tasks to have more in-flight MPI operations and to overlap the communication and computation phases.
In this evaluation there is one MPI rank per core for both the {\it Pure MPI} and {\it Interop} versions, so {\it Fork-Join} and {\it Sentinel} used on Gauss--Seidel would be equivalent to {\it Pure MPI}.

\begin{figure}
\centering
\includegraphics[width=1\linewidth]{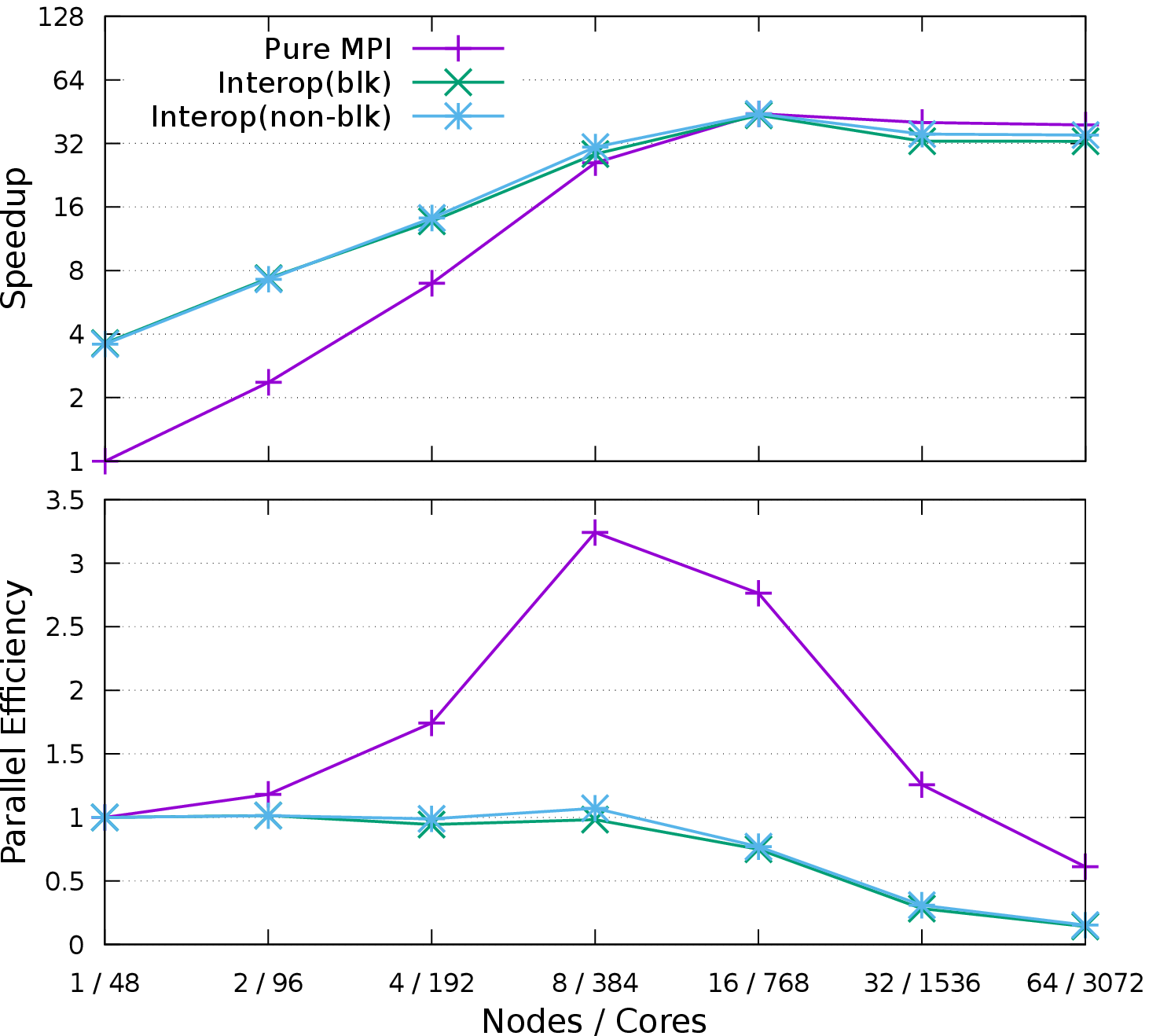}
\caption{Speedup and parallel efficiency of the IFSKer strong scaling (653K total gridpoints; 200 timesteps)}
\label{fig:ifsker_ss}
\end{figure}

We have executed the hybrid version with both the blocking and non-blocking interoperability modes of the TAMPI library previously explained: {\it Interop(blk)} and {\it Interop(non-blk)}, respectively.
Figure~\ref{fig:ifsker_ss} shows
the speed-up and parallel efficiency of the three versions on a
strong-scaling scenario. In the speed-up graph (upper part) we have used the performance of the {\it Pure MPI} version running on a single node as a baseline. For the parallel efficiency graph (lower part), each version uses as baseline its performance on a single node. 
The speed-up graph shows that, on a single node, the
performance of both {\it Interop} variants is 4x higher than that of the {\it Pure MPI} version.
These {\it Interop} variants scale linearly up to 16 nodes; after this point the problem size becomes too small.
Moreover, the {\it Interop(non-blk)} version performs slightly better than the {\it Interop(blk)} version, since the former prevents the communication tasks from being paused/resumed and it does not introduce any additional overhead.
It is worth noting that {\it Pure MPI} scales
superlinearly and with 16 nodes it reaches the {\it Interop}'s performance.
This effect is clearly reflected on the parallel efficiency graph (lower part of Figure~\ref{fig:ifsker_ss}). The parallel efficiency of {\it Pure MPI} grows until it reaches 3.2x at 8 nodes.   

%% file: tex/standardization.tex
\section{Standardization}
\label{sec:standardization}

In this section we will discuss the impact of the proposal in the OpenMP standard body.
Specifically, we describe the changes that need to be applied to the OpenMP standard and how they may affect its implementations.

\subsection{OpenMP}

The impact on the OpenMP standard could be measured according to two different
fronts: language and implementation. The impact on language affects OpenMP
programmers and the way they can interact with the programming model.
The impact on the implementation affects compiler--library providers and the way
the infrastructure should behave when executing the OpenMP program.

In terms of language the specification should include the eight API routines described
in Section~\ref{sec:apis} and provide the functionality to pause/resume
tasks, register/unregister polling services and bind/fulfill task external events.

In addition, this proposal should also impact on the specification's section concerning
task scheduling and, more specifically, the inclusion of new Task Scheduling Points (TSPs).
The call to the blocking service must be considered as a TSP allowing the implementation
to set aside the current task and start/resume the execution of any other task from the
ready task pool. The unblock service could also be included as a TSP allowing the scheduler
to continue with the execution of the current flow or resume the execution of the unblocked
task, but in this case it will be optional.

Finally, the Technical Report of OpenMP 5.0~\cite{openmp5.0} defines a task clause named {\it detach} which allows the
binding of a single event to a task so that the task does not fully complete until it has finished its execution
and the event has been fulfilled.
This functionality could be extended to allow the binding of more than one event to a specific task (as we proposed in
Section~\ref{sec:external_events_api}) since it could help users by facilitating the job of managing various
events produced by a task (e.g., the completion of different asynchronous operations).

%% file: tex/conclusion.tex
\section{Conclusion and Future Work}
In this paper we have presented the Task-Aware MPI library (TAMPI), which relies on the Pause/Resume and External Events APIs to support both blocking and non-blocking operations inside tasks.
TAMPI provides a new threading level called {\sf MPI\_\-TASK\_\-MULTIPLE} that enables task-aware blocking MPI operations. The library also provides two new functions {\sf TAMPI\_Iwait} and {\sf TAMPI\_Iwaitall} to efficiently support non-blocking MPI operations inside tasks.
This latter extension does not require any modification of the MPI standard.
In the evaluation section, we have demonstrated how TAMPI's enhanced support improves both programmability and performance of hybrid MPI+OmpSs applications.

On the one hand, using TAMPI's support for blocking MPI operations simplifies the porting of pure MPI codes to hybrid ones. In general it has very competitive performance but it can introduce some overheads when a large number of task are performing concurrently small MPI operations. In this case, a number of threads (and stacks) proportional to the number of in-flight MPI operations will be created, potentially degrading the performance of the application.
On the other hand, TAMPI's support for non-blocking MPI operations requires the use of a new call to register on-going MPI operations with the running task, but it completely avoids the creation of additional threads, associated context switches, and task scheduling points, therefore improving the performance in some scenarios.

As future work, we plan to study how TAMPI can be extended to support MPI RMA operations.
We also plan to explore the use of both APIs to improve the integration of task-based programming models with other blocking and non-blocking APIs.

\label{sec:conclusion}

%% file: tex/acknowledgments.tex
\section*{Acknowledgments}
This work has been developed with the support of the European Union H2020 Programme through both the INTERTWinE project (agreement No. 671602) and the Marie Sk\l{}odowska-Curie grant (agreement No. 749516); the Spanish Government through the Severo Ochoa Program (SEV-2015-0493); the Spanish Ministry of Science and Innovation (TIN2015-65316-P) and the Generalitat de Catalunya (2017-SGR-1414).